\documentclass{moriond}
\def\Journal#1#2#3#4{{#1} {\bf #2}, #3 (#4)}

\def\NIMA{{\em Nucl. Instrum. Methods} A}

\def\PLB{{\em Phys. Lett.}  B}
\def\PRL{\em Phys. Rev. Lett.}
\def\PRD{{\em Phys. Rev.} D}

\def\be{\begin{equation}}
\def\ee{\end{equation}}
\def\bea{\begin{eqnarray}}
\def\eea{\end{eqnarray}}

\newcommand{\Photo}{}

\begin{document}
\vspace*{4cm}
\title{CHARM AND $B$ TO CHARM DECAYS AT BELLE~II}

\author{ R.~MANFREDI, on behalf of the Belle~II Collaboration}

\address{University and INFN, Trieste}

\maketitle\abstracts{
We report on precise measurements of charmed hadron lifetimes from the Belle~II experiment and on the measurement of the angle $\gamma$ combining Belle and Belle~II data.\\
To measure charmed hadron lifetimes we use samples of electron-positron collisions collected from 2019 to 2021 near the $\Upsilon(4S)$ resonance,  corresponding to integrated luminosities of up to 207.2 ${\rm fb^{-1}}$. The results are the world's best determinations, indicating excellent early detector performance. 
We measure the angle $\gamma$ using a model-independent Dalitz plot analysis combining Belle and Belle~II data, with a total sample size of the order of 1 ${\rm ab^{-1}}$. This is the first combined analysis, and shows significant improvement with respect to previous $B$-factory determinations.
}

\section{Introduction}

The physics of charmed and beauty hadron decays is fundamental for the Belle~II physics program, with results expected to improve the precision of several sensitive tests of the Standard Model in the flavor sector. 
Belle~II is a particle detector designed to study 7-on-4 GeV electron-positron collisions at 10.58 GeV,  produced at very high luminosity by the SuperKEKB collider located at the KEK laboratory in Japan.~\cite{akai} The collision energy corresponds to the mass of the $\Upsilon(4S)$ resonance. The $\Upsilon(4S)$ decays almost exclusively into $B\bar{B}$ pairs with insufficient phase space to produce additional particles, which results in low backgrounds. The beam-energy asymmetry boosts the center of mass allowing measurable displacements of the $B$ decay vertexes from the interaction point, which are required for decay-time dependent measurements.  A  fraction of the data, called ``off-resonance", is collected at a collision energy $\approx60$ MeV below the $\Upsilon(4S)$ mass, to exclude the production of $B\bar{B}$ pairs and study non-$B$ backgrounds.\\
Belle~II consists of several subdetectors, arranged hermetically in a cylindrical geometry around the interaction point (IP). The innermost detector is a silicon tracker, based on pixel sensors for the first two layers (PXD) and on silicon strip sensors for the surrounding four layers (SVD).  The silicon tracker samples the trajectories of final-state charged particles at radii $1.4<r<13.5~{\rm cm}$ to reconstruct the decay position (vertex) of their long-lived parent particles, with a resolution of about 15 $\rm \mu m$. A large-radius wire drift chamber (CDC) measures charged-particle charges, momenta with 0.4\% resolution, and $dE/dx$ with $\sim$7\% resolution. A time-of-propagation Cherenkov detector and an aerogel ring-imaging Cherenkov detector surround the drift chamber and provide charged-particle identification (PID) information, allowing for separation of kaons from pions of up to 4 GeV/$c$ momentum, with 90\% efficiency and 5\% misidentification rate.  A CsI(Tl)-crystal electromagnetic calorimeter measures the energy of electrons and photons, with 1.6\%--4\% resolution. Layers of plastic scintillators and resistive-plate chambers alternated with iron plates provide muon and $K_{\rm L}^0$ reconstruction. Belle II started to collect data  in March 2019, aiming to accumulate a sample comparable in size to the ones from previous $B$ factories by summer 2022, and with the goal of collecting about forty times more data in the next decade.  
The Belle~II data used in this work have been collected up to the end of 2021, corresponding to integrated luminosities of up to 207.2 ${\rm fb^{-1}}$.~\cite{lumi} We report on precise measurements of charmed hadrons $D^0$, $D^+$, and $\Lambda_c^+$ lifetimes.  We also report on the measurement of the CKM angle $\gamma$, performed combining Belle~II data with the full Belle data set.

\section{Charmed lifetimes measurements}
\begin{figure}[h]
\begin{minipage}{0.33\linewidth}
\centerline{\includegraphics[width=\linewidth]{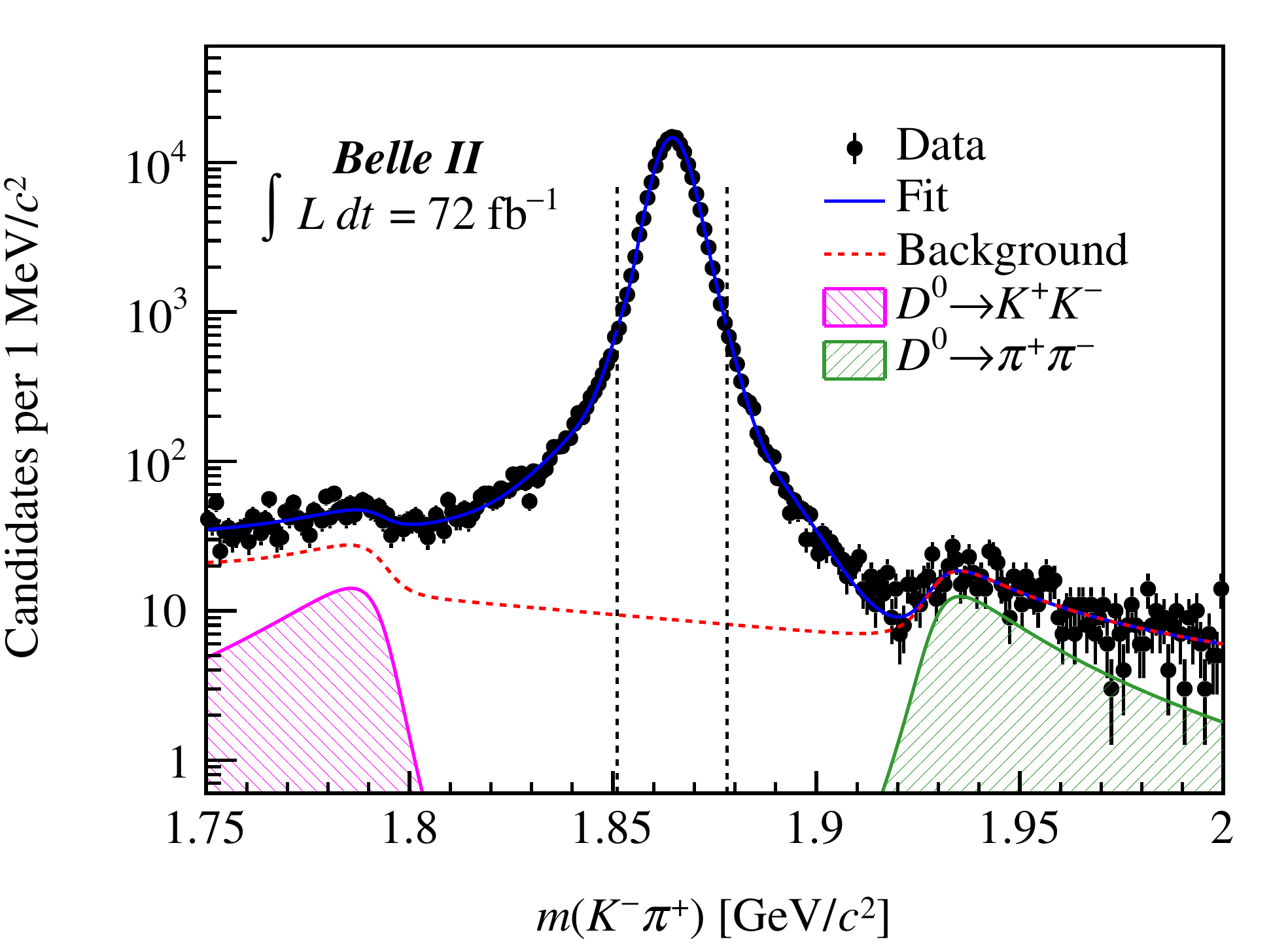}}
\end{minipage}
\hfill
\begin{minipage}{0.32\linewidth}
\centerline{\includegraphics[width=\linewidth]{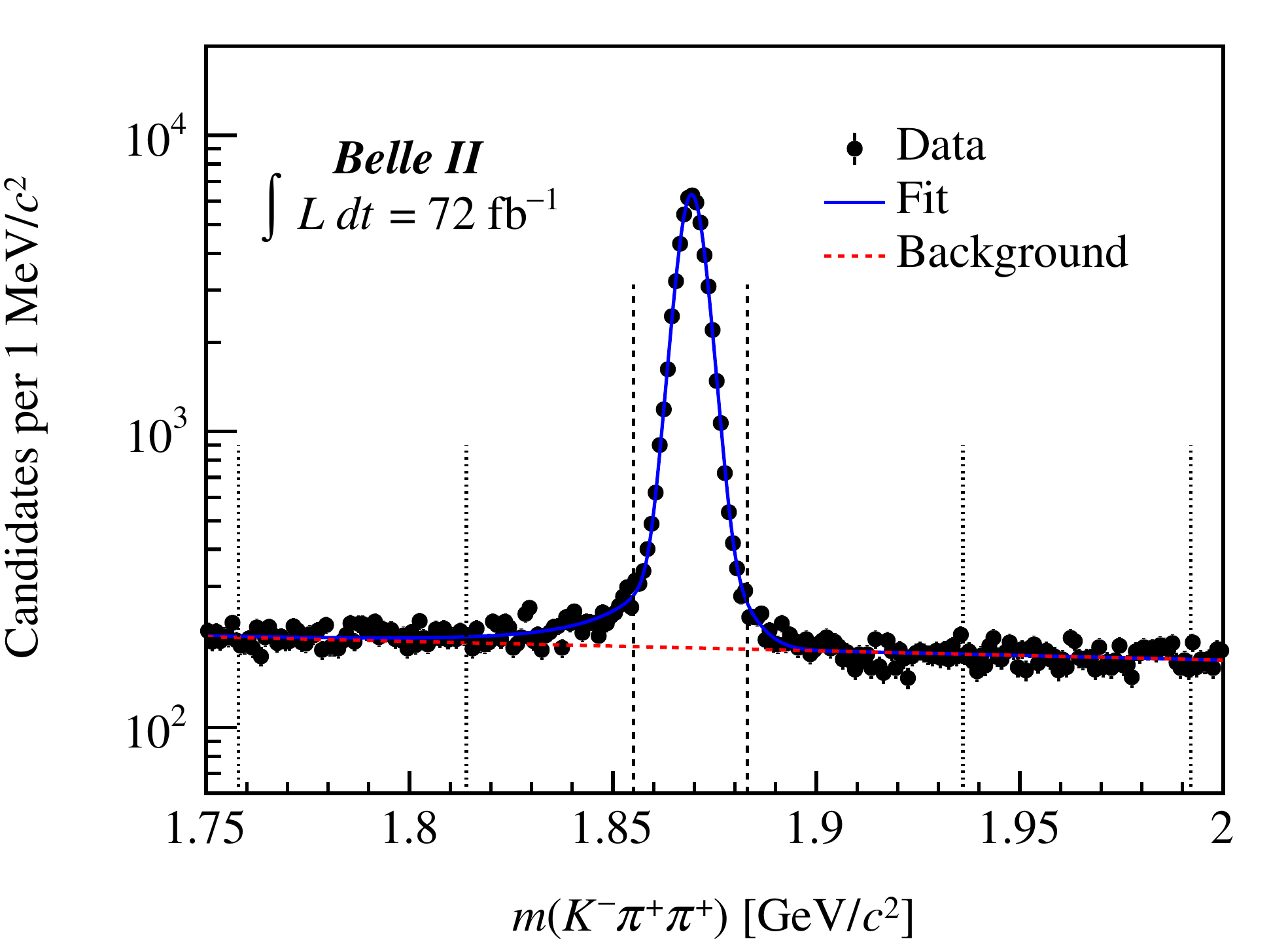}}
\end{minipage}
\hfill
\begin{minipage}{0.32\linewidth}
\centerline{\includegraphics[width=\linewidth]{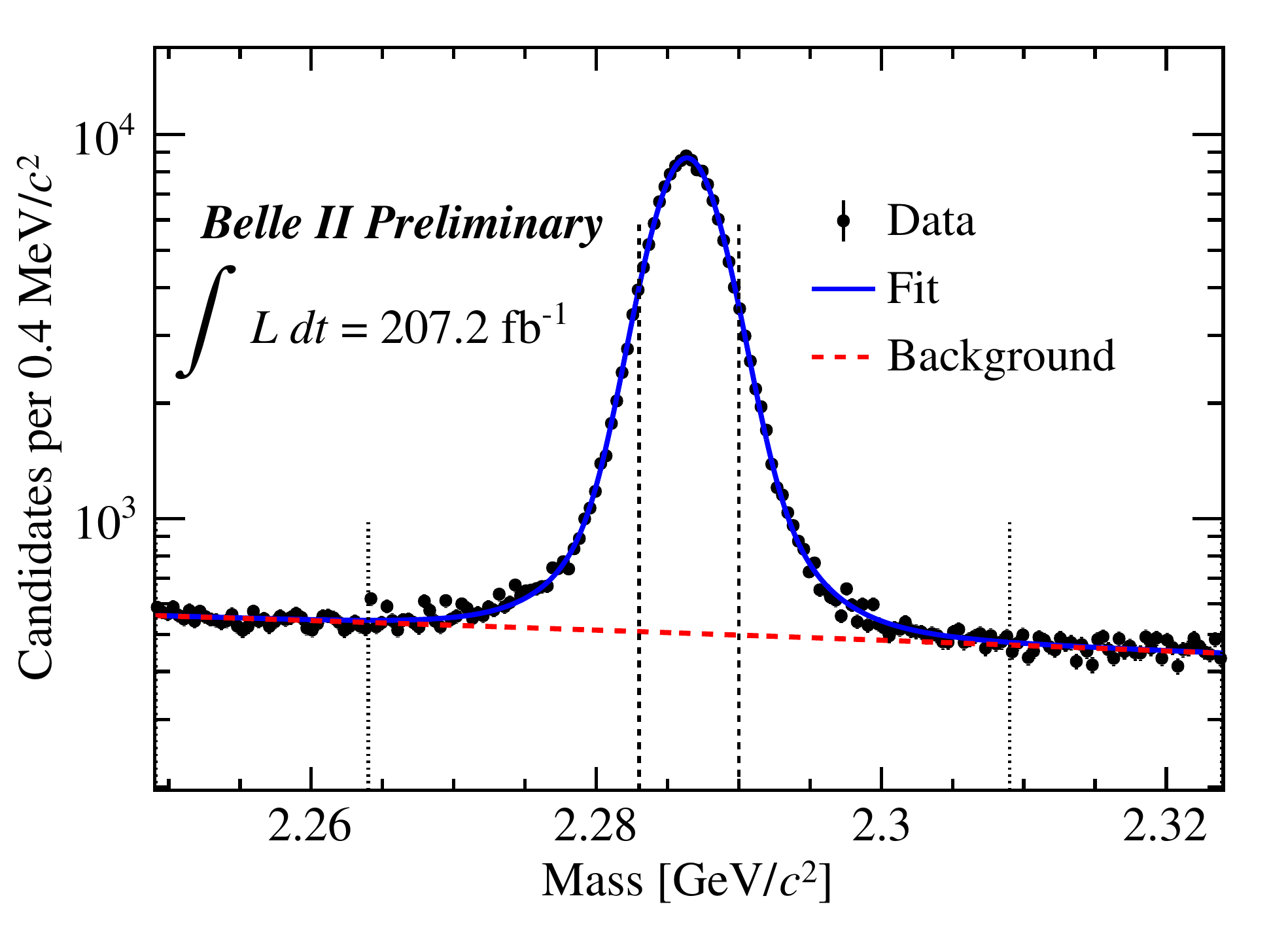}}
\end{minipage}
\caption[]{Invariant mass distributions of (left) $D^0\to K^-\pi^+$, (center) $D^+\to K^-\pi^+\pi^+$, and (right) $\Lambda_c^0\to pK^-\pi^+$ candidates with fit projections overlaid. The vertical dashed (dotted) lines indicate the signal region (sidebands).}
\label{fig:InvM}
\end{figure}

Theoretical predictions of charmed hadron lifetimes are challenging because they rely on the description of strong interactions at low energies. They are typically achieved using effective models, such as the heavy-quark expansion,~\cite{HQE1}~\cite{HQE2}~\cite{HQE3}~\cite{HQE4}~\cite{HQE5}~\cite{HQE6} that can be tested with precise lifetime measurements.
We reconstruct $D^{*+}\to D^0\left(\to K^-\pi^+\right)\pi^+$ and $D^{*0}\to D^+\left(\to K^-\pi^+\pi^+\right)\pi^0$ in 72 ${\rm fb^{-1}}$ of data,~\cite{Dpaper} and $\Lambda_c^+\to pK^-\pi^+$ in 207.2 ${\rm fb^{-1}}$ of data. The decay time of each candidate is measured from the projection of the displacement ($\vec{L}$) of the decay vertex from the production vertex on the momentum ($\vec{p}$) direction,  as $t=m\vec{L}\cdot\vec{p}/|\vec{p}|^2$, where $m$ is the mass of the charmed hadron. \\ 
Tracks are required to have at least one hit in SVD and 20 or 30 hits, depending on the decay channel, in the CDC. In the $D^0$ and $D^+$ lifetime measurements, a hit in the first layer of the PXD is also required. The lower-momentum pion in the $D^+$ lifetime measurement is required to have momentum greater than 0.35 GeV$/c$, while pions and protons in the $\Lambda_c$ lifetime measurement are required to have transverse momentum greater than 0.35 GeV$/c$ and 0.7 GeV$/c$ respectively. Low momentum $\pi^0$ are reconstructed from two photons as $\pi^0\to\gamma\gamma$. A global fit of the decay chain is performed,~\cite{treeFitter} and only candidates with $\chi^2$ greater than 0.01 are kept. The mass difference $\Delta m$ between $D^*$ and $D$ candidates is required to be 144.94 $<\Delta m<$ 145.90 MeV$/c^2$ for $D^0$ and 138 $<\Delta m<$ 143 MeV$/c^2$ for $D^+$ candidates. The IP is assumed to be the production vertex  of $D^*$ and $\Lambda_c^+$ candidates, therefore their momenta are required to be greater than 2.5 GeV$/c$ to reject the products of $B$ meson decays. Reconstructed candidates are further restricted in the invariant mass ranges of the $K^-\pi^+$, $K^-\pi^+\pi^+$ and $pK^-\pi^+$ respectively, after that fits of invariant mass distributions are performed to estimate the background fraction under the prominent signal peaks (Fig.~\ref{fig:InvM}).\\

\begin{figure}[h]
\begin{minipage}{0.33\linewidth}
\centerline{\includegraphics[width=\linewidth]{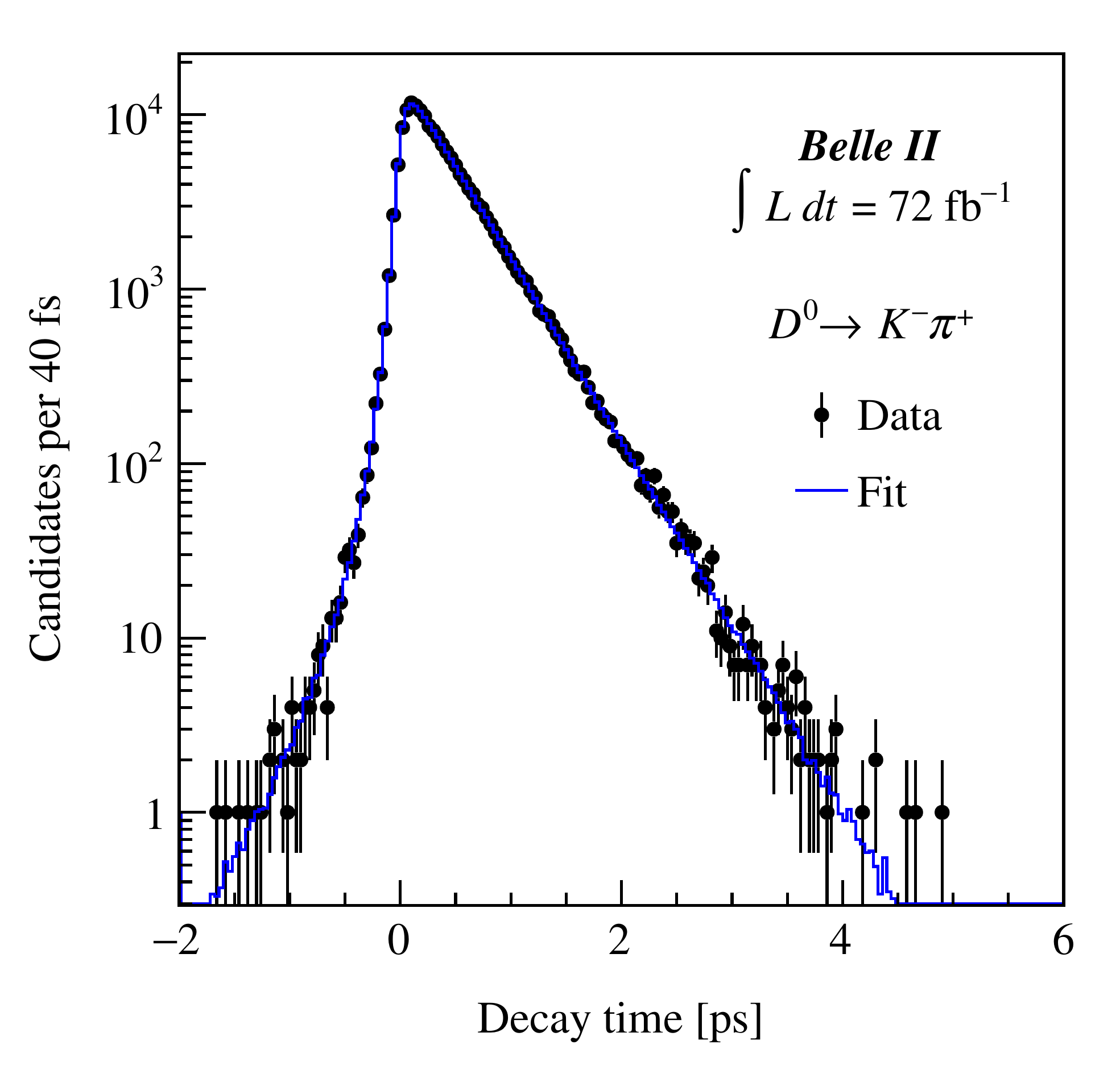}}
\end{minipage}
\hfill
\begin{minipage}{0.32\linewidth}
\centerline{\includegraphics[width=\linewidth]{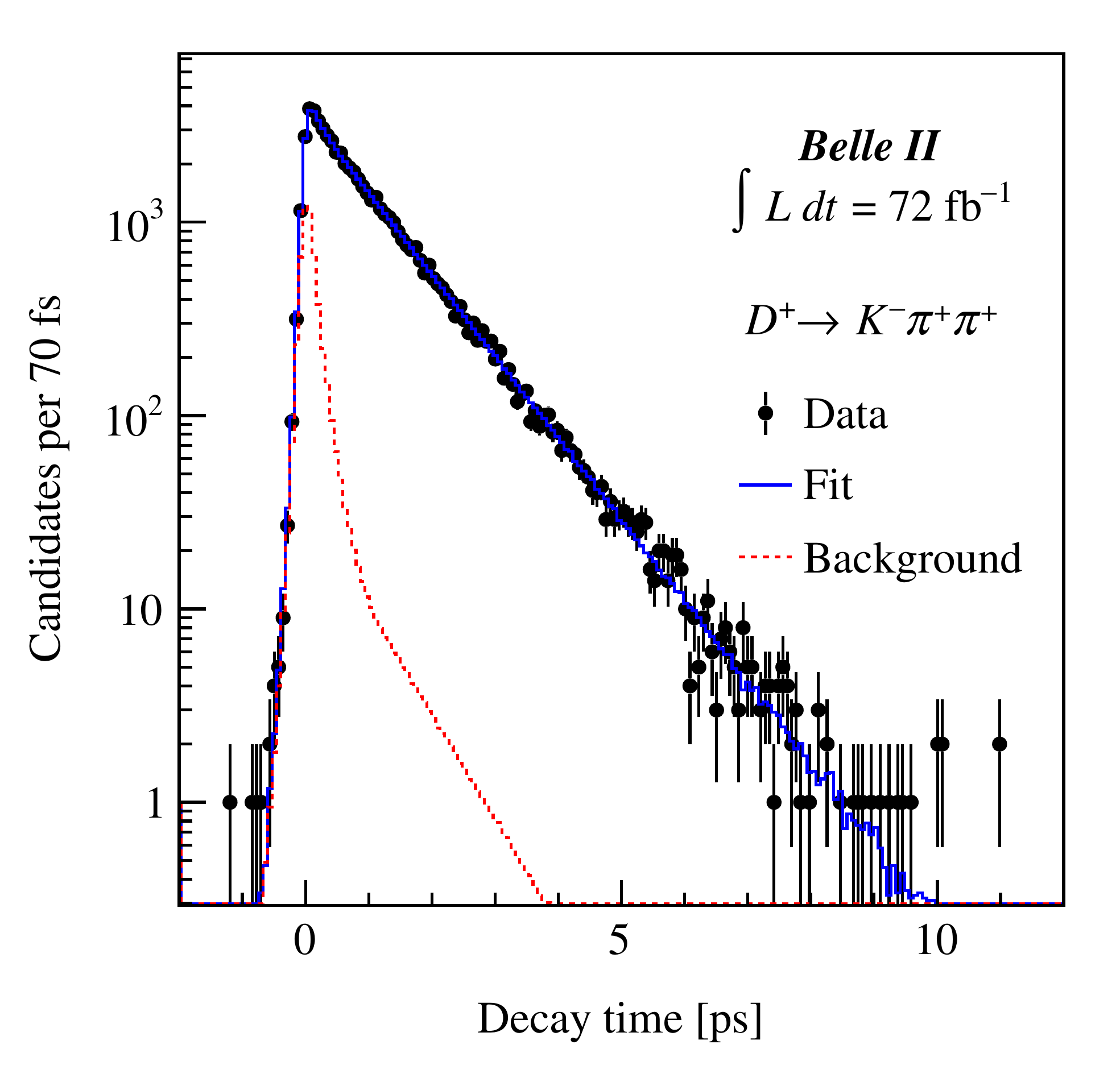}}
\end{minipage}
\hfill
\begin{minipage}{0.32\linewidth}
\centerline{\includegraphics[width=\linewidth]{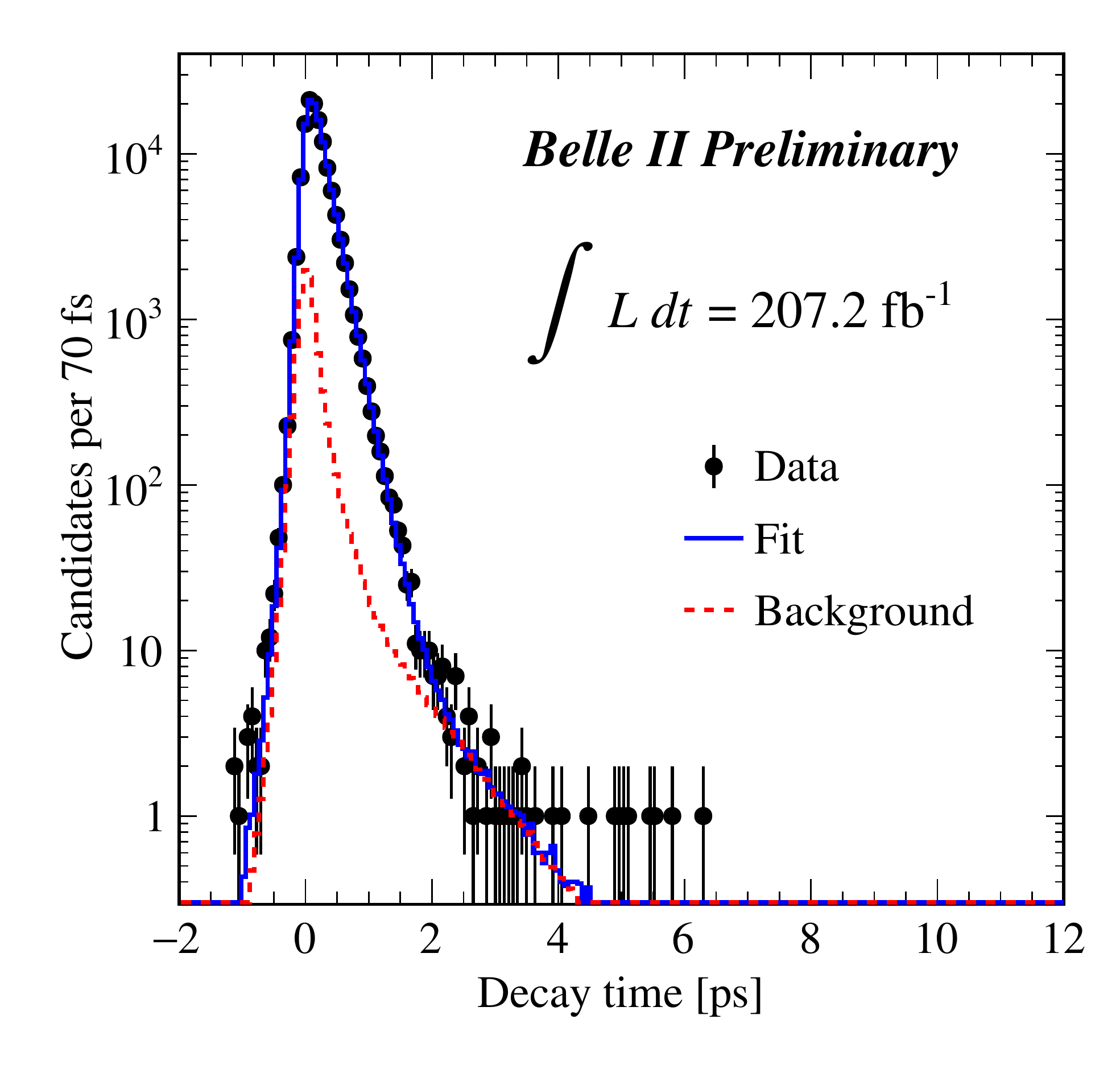}}
\end{minipage}
\caption[]{Decay time distributions of (left) $D^0\to K^-\pi^+$, (center) $D^+\to K^-\pi^+\pi^+$, and (right) $\Lambda_c^0\to pK^-\pi^+$ candidates in their respective signal regions with fit projections overlaid.}
\label{fig:Lifetime}
\end{figure}

Lifetimes are obtained with unbinned maximum-likelihood fits of the two-dimensional $\left(t,\sigma_t\right)$ distributions of candidates in the signal region, where $\sigma_t$ is the event-by-event uncertainty in $t$.  The signal probability density function (PDF) is the convolution of an exponential and a resolution function, made of one or two Gaussians that depend on $\sigma_t$, multiplied by a histogram template for $\sigma_t$, derived directly from data. All the parameters of the PDF are directly determined from the fit to the data, avoiding reliance on the simulation.
For the $D^0$ sample, the sub-percent background contamination is neglected; this assumption results in a small systematic uncertainty. Background contamination is accounted for in the $D^+$, and $\Lambda_c^+$ fits using sidebands in the invariant-mass distributions.  Background yields are constrained to the results of the invariant mass fits, and the PDFs are empirical models with all the parameters determined in the fit, which is simultaneously performed in the signal region and the sidebands.  Decay time distributions with fit results overlaid are shown in Fig.~\ref{fig:Lifetime}.\\

The main systematic uncertainty contribution comes from possible tracking detector misalignment, which can cause biases in the decay length determination. Such effects are estimated using simulations with realistic misalignments, extracted from alignment studies, and real data. Other source of systematic uncertainties, such as neglecting possible $t$-$\sigma_t$ correlation in resolution models or background distributions not well reproduced by the simulation, are quantified using simplified simulated experiments under different conditions.\\
The $\Lambda_c$ lifetime can be biased by contamination of $\Xi_c^{+,0}\to \Lambda_c^+\pi^{0,+}$ decays.  The $\Xi_c^0$ branching fraction is known,~\cite{XicLHCb} while the $\Xi_c^+$ is unobserved and only predicted to be about a factor two more abundant.~\cite{XicTh}  The systematic uncertainty associated with this is evaluated with simplified simulated experiments where realistic contributions of $\Xi_c$ decays are included. This evaluation results in an additional one-sided systematic uncertainty in this preliminary result. Studies are ongoing to suppress background of this kind, including reconstructing $\Xi_c^+$ candidates and vetoing their invariant mass range. All systematic uncertaintiy contributions are shown in Table \ref{tab:lifetimes}.\\
The measured lifetimes are
\begin{center}
$\tau\left(D^0\right) = 410.5 \pm 1.1 ({\rm stat.}) \pm 0.8 ({\rm syst}.) ~{\rm fs}$, 
\end{center}
\begin{center}
$\tau\left(D^+\right) = 1030.4 \pm 4.7 ({\rm stat.}) \pm 3.1 ({\rm syst}.) ~{\rm fs}$,
\end{center}
\begin{center}
$\tau\left(\Lambda_c^+\right) = 204.1 \pm 0.8 ({\rm stat.}) \pm 0.7 ({\rm syst.}) _{-1.4}  (\Xi_c)~{\rm fs}$.
\end{center}
These are the world's most precise determinations, and agree with current world averages.~\cite{PDG} Furthermore, these measurements validate the excellent vertexing performance of the Belle~II detector, which paves the way for future lifetime determinations and other time-dependent measurements.

\begin{table}[h]
\caption[]{Summary of the systematic uncertainties.}
\label{tab:lifetimes}
\vspace{0.4cm}
\begin{center}
\begin{tabular}{| l | c c c |}
\hline
Source & $\tau\left(D^0\right)$ $[{\rm fs}]$ & $\tau\left(D^+\right)$ $[{\rm fs}]$& $\tau\left(\Lambda_c^+\right)$ $[{\rm fs}]$\\
\hline
Resolution model & 0.16 & 0.39 & 0.46\\
Backgrounds & 0.24 & 2.52 & 0.20 \\
Detector alignment & 0.72 & 1.70 & 0.46\\
Momentum scale & ~0.19 & ~0.48 & ~0.09\\
\hline
Total & 0.8 & 3.1 & 0.7\\
\hline
$\Xi_c\to\Lambda_c\pi$ contamination & - - - & - - - & $-$1.4\\
\hline
\end{tabular}
\end{center}
\end{table}

\section{Combined measurement of $\gamma$ angle}
\begin{figure}[h]
\centering
{\includegraphics[width=0.4\linewidth]{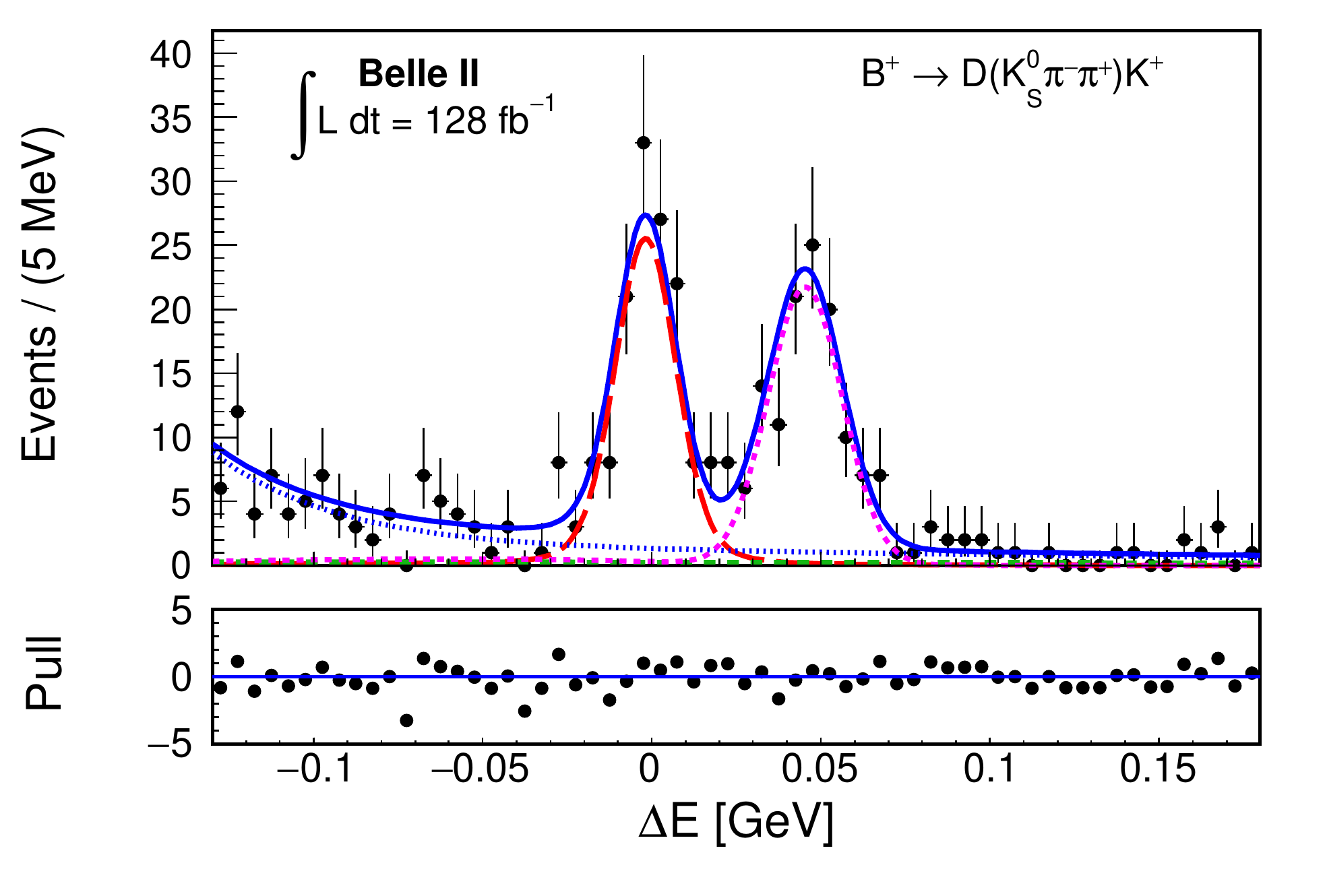}}
{\includegraphics[width=0.4\linewidth]{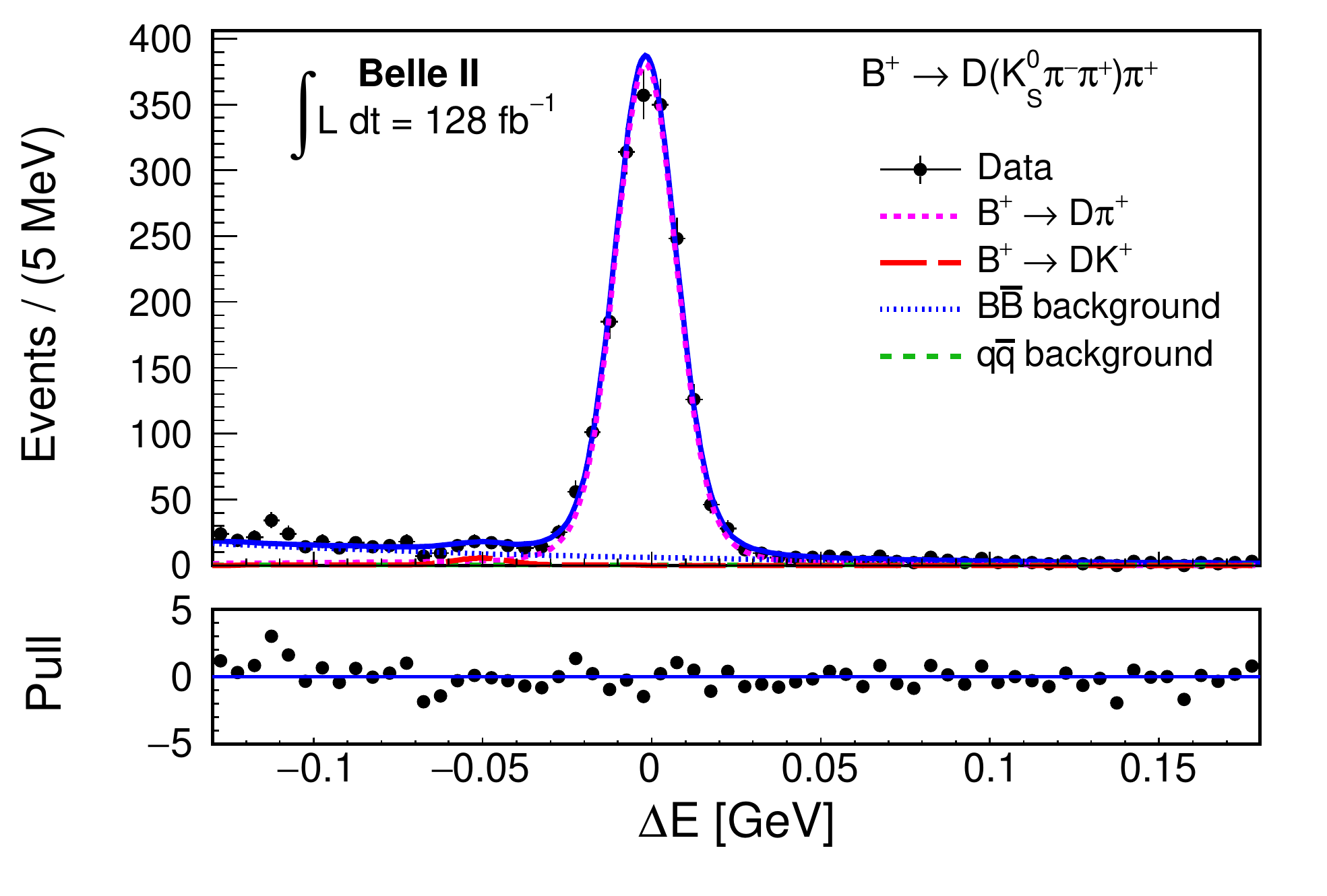}}
\caption[]{$\Delta E$ distributions of (left) $B^-\to D\left(\to K_S^0\pi^+\pi^-\right)K^-$ and (right) $B^-\to D\left(\to K_S^0\pi^+\pi^-\right)\pi^-$  candidates reconstructed in Belle~II data, with fit projections overlaid.}
\label{fig:gammaFitB2}
\end{figure}

\begin{figure}[h]
\centering
{\includegraphics[width=0.4\linewidth]{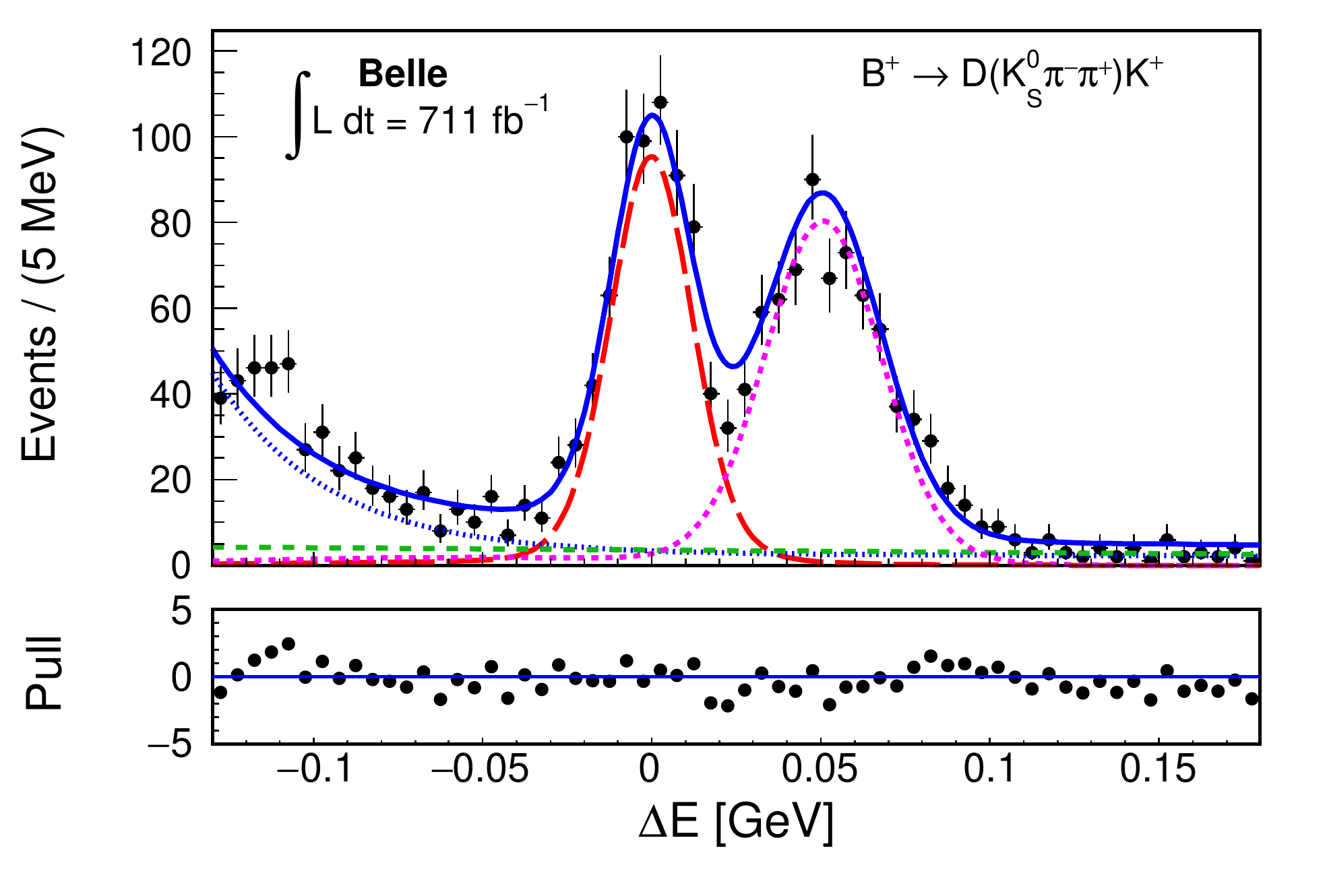}}
{\includegraphics[width=0.4\linewidth]{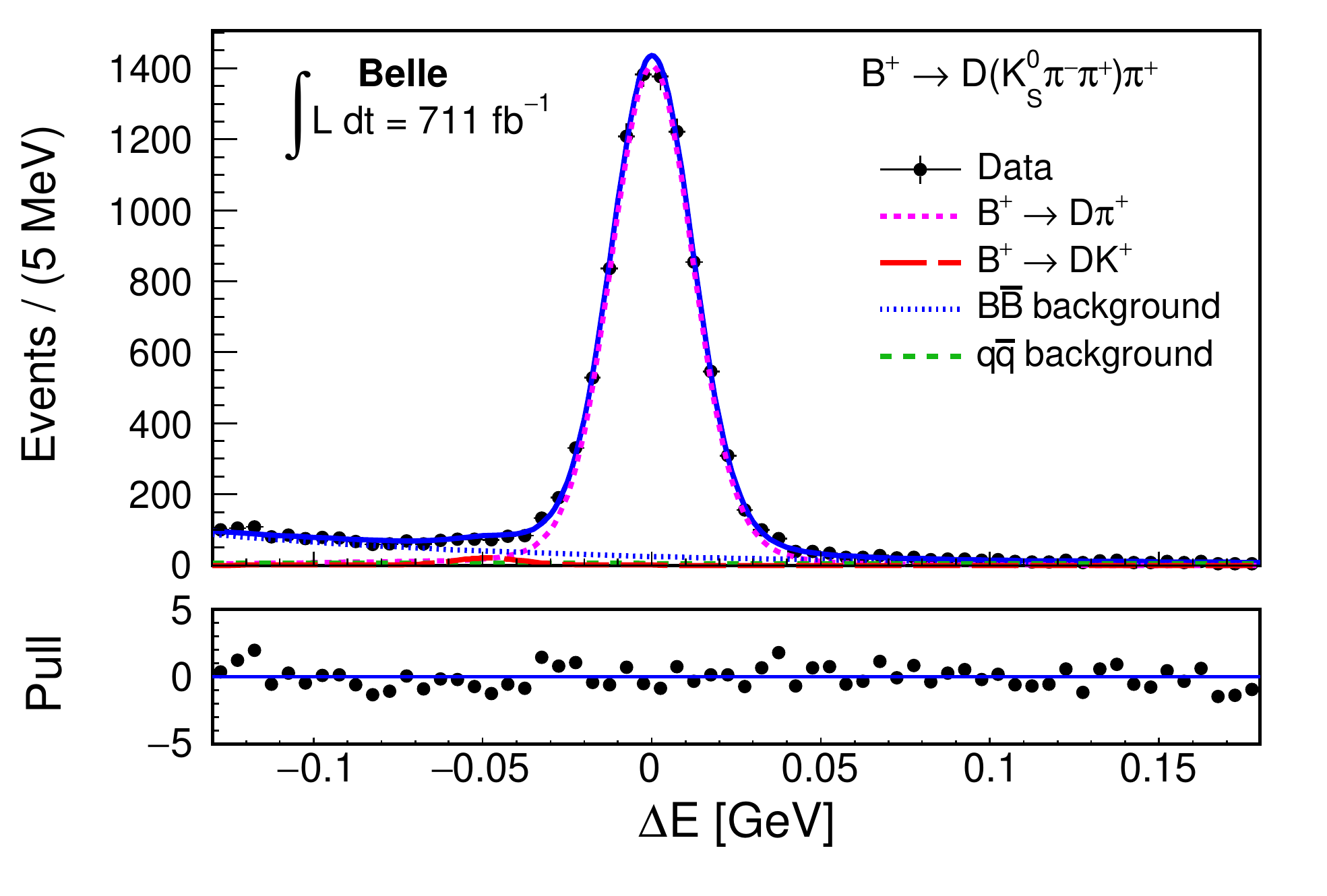}}
\caption[]{$\Delta E$ distributions of (left) $B^-\to D\left(\to K_S^0\pi^+\pi^-\right)K^-$ and (right) $B^-\to D\left(\to K_S^0\pi^+\pi^-\right)\pi^-$ candidates reconstructed in Belle data, with fit projections overlaid.}
\label{fig:gammaFitB1}
\end{figure}

Adding Belle~II data to the 711 ${\rm fb^{-1}}$ Belle data set allows impactful flavor physics measurements even with early data. The first of these combined measurements is the determination of the CKM angle $\gamma$.~\cite{gammaPaper}
Since it can be extracted using tree-level decays, where no non-SM physics is anticipated, the direct measurement of $\gamma$ provides a powerful gauge of the Standard Model when compared to indirect determinations based on other sides and angles of the Unitarity Triangle.
The most common channel to extract $\gamma$ is $B^{\pm}\to DK^{\pm}$, where $D$ indicates $D^0$ or $\bar{D}^0$ mesons decaying to the same final states $f$; the weak phase $\gamma$ enters in the interference of the favored $b\to c\bar{u}s$ and suppressed $b\to u\bar{c}s$ transitions $$ \frac{A\left(B^-\to \bar{D}^0K^-\right)}{A\left(B^-\to D^0K^-\right)} = r_B\exp^{i\left(\delta_B - \gamma\right)},$$ where $r_B$ and $\delta_B$ are the magnitude of the ratio and strong-phase difference between favored and suppressed amplitudes.\\
We reconstruct the decays $B^-\to D\left(\to K_S^0h^+h^-\right)h^-$, where $h$ is either a pion or a kaon. \mbox{$D\to K_S^0h^+h^-$} decays proceed through different intermediate resonances, resulting in ${\it CP}$ asymmetry parameters that vary over the phase space.  Following the BPGGSZ method,~\cite{BPGGSZL1}~\cite{BPGGSZL2}~\cite{BPGGSZL3} the $D$ Dalitz space is binned to achieve an optimal, model-independent determination of $\gamma$, eliminating systematic uncertainties due to the Dalitz model. The signal yield in each bin is $$ {\rm N}_i^{\pm} = {\rm h}_B^{\pm} \left[{\rm F}_i + r_B^2\bar{\rm F}_i +2\sqrt{{\rm F}_i\bar{\rm F}_i}\left(c_i x_{\pm} + s_i y_{\pm}\right)\right], $$ where $(x_{\pm},y_{\pm}) = r_B\left(\cos(\gamma + \delta_B), \sin(\gamma + \delta_B)\right)$.  Here ${\rm F}_i~(\bar{F}_i)$ is the fraction of $D^0~(\bar{D}^0)$ decaying in the $i$-th bin, while $c_i$ and $s_i$ are amplitude-averaged strong-phase differences between $D^0$ and $\bar{D}^0$ decays in the $i$-th bin, measured using pairs of $D$ mesons produced at $e^+e^-$ collider experiments operating at the $D\bar{D}$ production threshold.\cite{CLEO}~\cite{BESIII1}~\cite{BESIII2}\\
The selection is conceptually the same on the Belle and Belle~II data sets, even if criteria slightly differ.  Charged particles are reconstructed requiring a minimum distance of clocest approach to the interaction point in the transverse plane and along the beam axis. Tracks are identified as kaons or pions using PID information. $K_S^0$ candidates are reconstructed from two oppositely charged pion candidates, requiring the dipion invariant mass to be within $3\sigma$ from the known $K_S^0$ mass. The $D\to K_S^0h^+h^-$ candidates are then restricted to the known $D$ mass range $1.85 < m(K_S^0h^+h^-) < 1.88$ GeV$/c^2$.
$B$ candidates are formed combining $D$ and $K$ candidates. To further select the sample we use the variables $M_{\rm bc} = \sqrt{\left(\sqrt{s}/2\right)^2 - \left( \Sigma\vec{p}^*_i\right)^2}$ and $\Delta E = E^*_B - \left(\sqrt{s}/2\right)$, where $\sqrt{s}$ is the collision energy, $\vec{p}^*_i$ are the momenta of the reconstructed $B$ daughters and $E^*_B$ is the reconstructed energy of the $B$ meson, all in the center of mass frame. A kinematic fit is applied to the decay chain, constraining the $B$ daughters to come from the same vertex.\\
The dominant background comes from candidates reconstructed in $e^+e^-\to q\bar{q}$ ($q=u,d,s,c$) events. We suppress it combining many inputs on the event topology in a binary classifier based on boosted decision trees. This exploits the fact that these events produce particles collimated into back-to-back jets, while $e^+e^-\to\Upsilon(4S)\to B\bar{B}$ events have a more spherical shape.\\
The PID selection has a $O(10\%)$ misidentification rate, resulting in contamination of $B\to D\pi$ decays in the $B\to DK$ sample. To measure the PID efficiency and misidentification rate, we reconstruct also a disjoint sample of $B\to D\pi$ candidates. We extract these parameters fitting simultaneously the $B\to D\pi$ and $B\to DK$ samples in the Belle and Belle~II data sets, with an extended maximum-likelihood fit of the $\Delta E$ and transformed BDT output distributions. A projection of this fit on Belle~II (Belle) data is shown in Fig.~\ref{fig:gammaFitB2} (Fig.~\ref{fig:gammaFitB1}); the narrower $\Delta E$ peaks in the Belle~II data show the improvement in resolution with respect to Belle. Once these parameters are fixed, the same fit is performed simultaneously in all the Dalitz plot bins to measure the ${\it CP}$ asymmetry parameters $x_{\pm}$ and $y_{\pm}$. 

The values of $\gamma$, $r_B$ and $\delta_B$ are determined using the statistics package GammaCombo,~\cite{statPack} choosing the solution with $0^{\circ}<\gamma<180^{\circ}$ as suggested by other measurements.~\cite{HFLAV} The results are
\begin{center}
$\gamma = \left(78.4 \pm 11.4 ({\rm stat}) \pm 0.5 ({\rm syst}) \pm 1.0\ ({\rm ext})\right)^{\circ}$, 
\end{center}
\begin{center}
$r_B = 0.129 \pm 0.024 ({\rm stat}) \pm 0.001 ({\rm syst}) \pm 0.002\ ({\rm ext})$,
\end{center}
\begin{center}
$\delta_B = \left(124.8 \pm 12.9 ({\rm stat}) \pm 0.5 ({\rm syst}) \pm 1.7\ ({\rm ext})\right)^{\circ}$,
\end{center}
where the third uncertainty refers to the external inputs for $c_i$ and $s_i$ values.\\
Results are compatible with and more precise than the previous Belle result.~\cite{gammaBelle} Systematic uncertainties are reduced because of the improved background suppression and the use of $B\to D\pi$ sample in the fit, reducing the reliance on simulation. The external uncertainty is strongly reduced because of the new inputs from the BESIII collaboration. The net improvement is equivalent to doubling the sample size, even if Belle~II data set only corresponds to about 20\% of the Belle one.
The precision is limited by the sample size,  with extrapolations indicating that future analysis of the same decay mode will reach $4^{\circ}$ precision on the $B^-\to D\left[\to K_S^0h^+h^-\right]h^-$ mode alone with approximately 10 ab$^{-1}$,  reducing to $3^{\circ}$ when more $D$ final states will be included.

\newpage
\section{Summary}
We report on the world's best measurements of $D^0$, $D^+$ and $\Lambda_c^+$ lifetimes, and on the first combination of Belle and Belle~II data, to measure the CKM angle $\gamma$.\\
Charmed hadron lifetimes are measured on samples of 72~${\rm fb^{-1}}$ and 207.2~${\rm fb^{-1}}$ of Belle~II data.  These are the world's best determinations, and prove the excellent vertexing performance of the detector, and the capability of providing competitive results already with early data.\\
The measurement of the CKM angle $\gamma$ is the first combination of Belle and Belle~II data. Uncertainties are dominated by the sample size. The precision is significantly improved with respect to the previous Belle determination, showing good understanding of the detector and of the analysis procedure.

\end{document}